# Electron magnetic resonance studies of nanomanganite $Nd_{0.67}Sr_{0.33}MnO_3$

**S.S.Rao[1] , Subhasis Sarangi[1], G. Venkataiah[2], Venu Gopala Reddy[2] and S.V.Bhat[1].**

**[1]Department of Physics, Indian Institute of Science,Bangalore, India, 560012.**

**[2]Department of Physics, Osmania University, Hyderabad, India.**

$Nd_{0.67}Sr_{0.33}MnO_3$ (NSMO) nano particles with the grain size of about 30 nm are prepared by sol-gel method. These nanopowders are annealed at four different temperatures viz 800° C, 900° C, 1000° C and 1100° C to study the effect of particle size on magnetic, transport and electron magnetic resonance spectral parameters. The samples are characterized by XRD, SEM, EDAX and TEM. The a.c susceptibility experiments show that as the particle size increases the ferromagnetic to paramagnetic transition temperature ($T_c$) decreases. The metal-insulator transition temperature also changes with the particles size as revealed by resistivity measurements. Electron magnetic resonance (EMR) spectra of the nano powders are recorded from room temperature down to 4K using an X- band ESR spectrometer. EMR spectra could be fitted using two broad-Gaussian lineshapes below $T_c$ and suggested the ferromagnetic nature of the sample.Above $T_c$ a single Lorentzian fits the signals as expected for a paramagnetic sample. The EMR spectral parameters are found to be different from the bulk(polycrystalline) sample data .The spectral parameters show variation with the particle size. The presence of the two signals in the ferromagnetic phase is attributed to core and shell regions of the nano particles. We could estimate the shell thickness from the EMR intensity data as 0.7-1nm which agrees well with the other measurements.

**Introduction:** The magnetic and transport properties of rare-earth manganites of the general formula $Ln_{1-x} A_x MnO_3$ (Ln = rare earth element, A = alkaline earth element) have been systematically investigated for single-crystals , thin films and polycrystalline samples due to their importance for fundamental research and potential applications [1-5]. Most attention has been focussed on the the prototypical compound with x near 0.33, which exhibits an optimal ferromagnetism and magnetoresistance. Recently it has been shown that the transport properties of perovskite manganites not only depend up on the method of preparation, but also on the particle size [6-10]. Some of the salient features exhibited by the manganite systems when we reduce the particle size are i)showing the core-shell behaviour, ii)a decrease and broadening of the ferromagnetic transition temperature $T_c$, iii) showing super paramagnetic behaviour and intergranualar magnetoresistance (IMR), iv) tuning of CMR behavoiur and the decrease in the magnetization in comparison with the single crystal and bulk polycrystalline samples, v) a higher value of magnetoresistance compared to the bulk samples especially at low temperature etc..

There are some reports of enhancement of $T_c$ in the nanostructered manganites when compared to it's bulk counterpart[7,10,24]. NSMO shows the interesting and rich phase diagram across it's composition and temperature. It shows the para to ferromagnetic transition at the temperature ~ 200K. In this study, we have shown that the ferromagnetic transition temperature $T_c$ increases with the decrease of particle size and this enhancement is very important from the technological point of view. It is also seen that the metal-insulator transition temperature ($T_p$)increases with the decrease of particle size. There are very few reports of Electron Magnetic Resonance (EMR) studies of nanomanganites and are mainly focussed in the paramagnetic region[28]. We have done an extensive EMR studies on NSMO of different particle sizes across it's magnetic phase transition. The important finding of these EMR studies is that probing the core and shell regions of different magnetic nature in a nanoparticle experimentally by analyzing the spectral parameters with the temperature, which supports the existing theortical and experimental models [11-12 ].

**Experimental details:** Single phase, nanocrystalline samples of NSMO were prepared by the sol-gel method. Stochiometric amounts of $Nd_2O_3$ , $SrCo_3$ and $MnO_2$ were used as starting materials. These ingredients are converted into their corresponding nitrates by dissolving them in dilute nitric acid. The nitrates were mixed in the solution, citric acid was added to it, and the resulting solution is slowly evopourated to get a pink colored gel. The gel when decomposed at about 250K-300K, resulted in a highly porous black powder. Th e resulting powder was separated into four parts and annealed for about 7 hours at 800°C, 900°C, 1000°C and 1100°C to get different particle sizes and are named as NSMO-8, NSMO-9, NSMO-10, NSMO-11. The samples were characterized by x-ray diffraction using a PHILIPS XPERT based diffractometer and the particle size was determined independently from x-ray and transmission electron microscopy (TEM) measurements. Resistivity measurements were done using the standard four probe technique from 77K-350K in both in the presence and the absence of magnetic field. A home-made AC susceptibility set up was used to measure the magnetic phase transitions of all the four samples. The EMR experiments were carried out using a Bruker ER 200D ESR spectrometer on dispersed (in paraffin wax) nanoparticles as the temperature ranges from 4K to 310K. DPPH was used as a field marker to measure the g-value accurately.

**Results and Discussion:**

**x-ray diffraction:**

The X-ray diffraction data confirms the orthorhombic structure of all the samples. In order to find the sintering effects on the crystallite size, which has an influence on electrical and magnetic transport properties, the average crystallite size <S> of all the materials has been calculated using the relation [13],

$\langle S \rangle = K\lambda / \beta \cos(\theta)$ ………………………….(1)

where K is a constant depending on the grain shape. (K=0.89),

$\lambda$ is wavelength of CuK$\alpha$ radiation, ($\lambda$ = 1.541A$^o$)

$\beta$ is full width at half maxima of XRD peak.

and are found to vary from 15- 30 nm as the sintering temperature varies from 800$^o$C to 1100$^o$C as shown in table 1. With a view to study how the crystallite size varies with varying sintering temperature, the microstructure of all the materials was studied and the figures are given Fig. 1. It can be seen from the figures that there is a clear variation in the crystallite size as the sintering temperature is increased from 800 to 1100 $^0$ C. Full profile fitting refinements of the powder diffraction patterns of all the four samples were per formed using the program FULLPROF, based on the Rietveld method. We obtained the best fit for the orthorhombic (space group $P_{BNM}$) structure for all the four samples. The unit cell parameters fo are a = 5. 45 A$^o$, b = 5.43 A$^o$ and c = 7.71 A$^o$ for NSMO-8 sample and the corresponding parameters for it's bulk counter part are a = 5.46 A$^o$, b = 5.45 A$^o$ and c = 7.73 A$^o$ [29]. From the unit cell parameters, it is observed that there is a slight contraction in unit cell volume.

**b. Transmission Electron Microscopy (TEM):**

TEM micrographs are shown for the two samples NSMO-8 and NSMO-11 in the fig 2. It clearly shows that the grain size has been increased from 15nm-30nm (average) as the sintering temperature increase from 800$^o$C to 1100$^o$C.

**AC susceptibility measurements:**

AC susceptibility measurements were carried out using a home-made susceptibility set up from 80K-350K. The ferromagnetic phase transition temperatures increase with the decrease of particle size as presented in the table 1.

| nple code | ompositional Formula | ring Temp. ($^o$C) | $T_C$ ΔT ($T_C$ ~ $T_P$) ------------------------------- ( Degree Kelvin ) | | | S m) | R% |
|---|---|---|---|---|---|---|---|
| SMO-8 | $d_{0.67}Sr_{0.33}MnO_3$ | 800 | 15 | 260 | 45 | 5 | 5 |
| SMO-9 | $d_{0.67}Sr_{0.33}MnO_3$ | 900 | 25 | 258 | 33 | 0 | 7 |
| SMO-10 | $d_{0.67}Sr_{0.33}MnO_3$ | 1000 | 40 | 253 | 13 | 5 | 4 |
| SMO-11 | $d_{0.67}Sr_{0.33}MnO_3$ | 1100 | 45 | 249 | 4 | 0 | 5 |

**TABLE: - 1 - The experimental data of NSMO materials**

The enhancement in $T_c$ may be explained in the following way. Rietveld refinement analysis of XRD pattern shows the slight reduction in the unit cell volume (~ 1%). The enhancement may be due to the "unit cell volume contraction" and the reduction in the unit cell anisotropy parameter [24]. This causes the decrease and increase of bond length and bond angle respectively, which enhances the bandwidth, transfer ingral which makes the electron to hop easily and thereby shows the enhancement in $T_c$.

**Electrical transport and magnetoresistance:** The conductivity behaviour is explained both in the metallic region and in the insulating region.

**1. Ferromagnetic metallic region ($T < T_P$):**

In spite of over a decade of intense work on CMR materials, the temperature dependent resistivity data at low temperature and relative strengths of the different scattering mechanisms originating from different contributions of the manganites system is not yet understood thoroughly [14]. In order to understand the nature of the conduction mechanism at low temperature ($T<T_p$), we tried to fit the resistivity data with three empirical equations derived by different authors [12,15,16].

$$\rho = \rho_0+\rho_2 T^2 \qquad \text{………………(3)}$$

$$\rho = \rho_0+\rho_{2.5} T^{2.5} \qquad \text{…...………(4)}$$

$$\rho = \rho_0+\rho_2 T^2 +\rho_{4.5} T^{4.5} \qquad \text{…. ... .. ....…….(5)}$$

In the above equations, $\rho_0$ represents the resistivity due to grain boundary effects. $\rho_2 T^2$ term in equations -3 and 5 indicates the resistivity due to electron – electron scattering process, and is generally dominant upto 100K. On the other hand, the term $\rho_{2.5} T^{2.5}$ represents the resistivity due to electron – magnon scattering process in ferromagnetic phase. Finally, the term $\rho_{4.5} T^{4.5}$ indicates the resistivity due to electron – magnon scattering process in ferromagnetic region, which may be likely to arise due to spin wave scattering process.

The experimental data of four samples were fitted to the above three equations and the quality of these fittings, in general is evaluated by comparing the square of linear correlation coefficient ($R^2$), obtained for each equation. Therefore, $R^2$ values of all the samples for each equation were calculated and are listed in Table 2.

| ample code | $= \rho_0+\rho_2T^2$ | $= \rho_0+\rho_{2.5}T^{2.5}$ | $= \rho_0+\rho_2T^2 +\rho_{4.5}T^{4.5}$ |
|---|---|---|---|
| NSMO-8 | 0.9910 | 0.9894 | 0.9993 |
| NSMO-9 | 0.9977 | 0.9945 | 0.9992 |
| NSMO-10 | 0.9961 | 0.9946 | 0.9993 |
| NSMO-11 | 0.9931 | 0.9946 | 0.9993 |

**Table: 2. The square of linear correlation coefficient ($R^2$) values in ferromagnetic region**

It is interesting to note that $R^2$ values are found to be as high as 99.9 % when the data are fitted to equation-5[15-16]. The best fit parameters obtained for all the materials at 0T and 7 T fields are given in Table 2. The $\rho_0$ and $\rho_2$ values found to decrease with increasing sintering temperature and in fact, the observed $\rho_o$ values are larger than those obtained in the case of single crystals [17]. It means that both these parameters are decreasing with increasing grain size, which may be an evidence for the decrease of scattering processes due to the enlargement of the grains of the material. Thus the increasing grain size may decrease the grain boundary region and net grain boundary scattering term as well as electron-electron scattering term will decrease. The last term $\rho_{4.5}$ is also found to decrease with increasing sintering temperature and the observed behavior may be due to partial alignment of the spins which results in lowering their fluctuations [15]. Therefore grain boundary plays a dominant role in the conduction process and it acts as the region of enhanced scattering center for conduction electron [13].

Further, all the three parameters $\rho_o$, $\rho_2$ and $\rho_{4.5}$ in the presence of magnetic field, have also been computed for all the materials and are given in Table 3. All the three parameters are found to decrease with increasing magnetic field. The observed behavior may be explained as follows- when magnetic field increases, the domain gets enlarged so that the value of $\rho_0$ decreases, while the reduction in $\rho_2$ and $\rho_{4.5}$ could be attributed to the decrease of electron spin fluctuations in the presence of magnetic field.

| ample Code | $\rho_0$ ($\Omega$cm) | | $\rho_2$ ($\Omega$cm $K^{-2}$ ) | | $\rho_{4.5}$($\Omega$cm $K^{-4.5}$) | |
|---|---|---|---|---|---|---|
| | T | T | 0T | 7T | 0T | 7T |
| SMO-8 | 75 | 09 | $.00\times10^{-4}$ | $00\times10^{-4}$ | $.18\times10^{-10}$ | $13\times10^{-10}$ |
| SMO-9 | 44 | 81 | $40\times10^{-4}$ | $20\times10^{-4}$ | $.97\times10^{-11}$ | $39\times10^{-11}$ |
| MO-10 | 09 | 33 | $81\times10^{-4}$ | $80\times10^{-4}$ | $.94\times10^{-11}$ | $64\times10^{-12}$ |
| MO-11 | 88 | 89 | $.7\times10^{-4}$ | $50\times10^{-4}$ | $.08\times10^{-11}$ | $15\times10^{-12}$ |

**TABLE: 3. The best fit parameters obtained from experimental resistivity data in the ferromagnetic metallic region .**

Intrinsically, The scattering effects are suppressed from various contributions because of the parallel configuration of the spins present in the domain [13], and hence all the contributing parameters to the resistivity viz; $\rho_{o,}$ $\rho_2$ and $\rho_{4.5}$ may decrease with the application of the magnetic field [18].

**2 . Paramagnetic insulating region**

In order to explain the high temperature (T > Tp) resistivity data, two models viz; variable ranging hopping model ($Tp<T<\theta_D/2$) and the small polaron hopping ($T>\theta_D/2$) are generally used.

**a). Variable Range Hopping (VRH) model.** ($Tp < T < \theta_D /2$)

The Mott's equation for VRH mechanism [18] given by,

$$\sigma = \sigma_0 \exp (- T_0 / T)^{1/4} \quad \ldots\ldots\ldots\ldots\ldots\ldots(6)$$

where $\sigma_0$ is pre factor, was used to explain the conductivity data at temperatures - $Tp < T < \theta_D /2$, where $\theta_D$ is Debye temperature. The resistivity data, above $T_P$ is fitted to equation 6 by plotting ln ($\sigma$) vs $T^{-1/4}$ and from the best fits $\theta_D / 2$ values are estimated. Here $\theta_D / 2$ is defined as the temperature at which deviation from linearity occurs in the temperature region above $T_P$. The $\theta_D$ values are found to increase systematically with sintering temperature and are given in Table 4 and these are in fact very close to the reported ones [14]. Further $T_o$ values for each sample were calculated from slopes of ln($\sigma$) vs $T^{-1/4}$ plot. Finally, using the $T_o$ values and the equation,

$$T_0 = 16\alpha^3/K_BN(E_F) \quad \ldots\ldots\ldots\ldots\ldots\ldots\ldots.(7)$$

N ($E_F$), the density of states at the Fermi level for each material was also obtained. Here, the value of $\alpha$ =2.22nm$^{-1}$ has been used for calculations which was estimated and reported by Viret et al. [19].

| ple code | (K) | $E_P$ ( meV) | | $T_o$ ($10^6$ K) | | N($E_F$) (eV$^{-1}$ cm$^{-3}$) | |
|---|---|---|---|---|---|---|---|
| | | =0T | =7T | =0T | =7T | B=0T | B=7T |
| SMO-8 | 0.4 | 0.99 | 5.99 | .87 | .49 | 24×10$^{20}$ | 14×10$^{21}$ |
| SMO-9 | 0.8 | 0.92 | 7.82 | .56 | .31 | 93×10$^{20}$ | 51×10$^{21}$ |
| SMO-10 | 0.5 | 5.95 | 9.28 | .05 | .20 | .25×10$^{20}$ | 80×10$^{21}$ |
| SMO-11 | 1.1 | 8.83 | 3.84 | .63 | .13 | .96×10$^{20}$ | .60×10$^{21}$ |

**TABLE: 4. The best fit parameters obtained from the experimental resistivity data in the paramagnetic insulating region.**

All the estimated parameters are given in Table 3 and are found to be in agreement with those reported in the literature for other manganite materials [13,15,20]. It can also be seen from the table that $T_0$ values are found to decrease enormously and continuously with increasing sintering temperature as well as with increasing magnetic field. Further, for a given material $T_0$ values are found to be lesser in the presence of the field and this can be due to suppression of the magnetic domain scattering with the application of the field [13], consequently the values of the density of states are found to increase continuously with increasing sintering temperature. In fact a similar trend has been observed even in the presence of magnetic field also.

**b). Small polaron hopping model.** ($T > \theta_D/2$)

As mentioned earlier, the conduction mechanism of these materials at high temperatures ($T > \theta_D/2$) is governed by small polarons and could be due to either adiabatic or non-adiabatic approximations [21]. The temperature dependence of the electrical resistivity arising out of adiabatic and non-adiabatic approximations are given as:

$$\rho = \rho_\alpha T \exp (E_P/K_BT) \qquad \text{(adiabatic)}\ldots\ldots\ldots\ldots\ldots\ldots\ldots\ldots(8)$$

$$\rho = \rho_\alpha T^{3/2} \exp (E_P/K_BT) \qquad \text{(non-adiabatic)}\ldots\ldots\ldots\ldots\ldots\ldots(9)$$

where $\rho_\alpha$ is the residual resistivity and $E_P$ is the activation energy.

According to Jung et al. [22], higher values (two to three orders than those of usual oxide semiconductors) of N ($E_F$) in the manganite system could be due to their high value of conductivity. These higher values of N ($E_F$) are clear signatures of the applicability of the adiabatic hopping mechanism. Based on this fact, the adiabatic small polaron hopping model rather than the non-adiabatic small polaron hopping model can be used in the present investigation.

The $E_P$ obtained from the slopes and intercepts of the ln(ρ/T) vs1/T plots are given in Table 4 and it can be seen from the table that the activation energy values are found to decrease continuously with increasing sintering temperature. $E_P$ values are decreasing with increasing grain size both in presence and in absence of the field and the observed behavior may be due to the fact that with increasing grain size interconnectivity between grains increases, which facilitates the hopping of the electron to the neighboring sites [13]. Further, a similar decrease in the values of $E_p$ even in the presence of magnetic field has also been observed and the observed behavior may be due to the decrease in the values of charge localization when magnetic field is applied [23].

**Electron Magnetic Resonance studies:** Electron Magnetic Resonance (EMR) is a powerful microscopic technique to probe the complex spin dynamics, magnetic phase transitions and phase seperations [27,30] in rare earth manganites**.** The NSMO nano crystals are dispersed in paraffin wax to isolate the powder particles both electrically and magnetically. To determine the accurate value of the "g" factor , a speck of DPPH (diphenyl-picryl-hydrazyl) was used as a field marker in all the experiments. Bruker ER 200D ESR spectrometer has been used to record the signals at the temperature ranging between 4.2K-305K. In the ferromagnetic phase , EMR signals fit to two Gaussians indicating the presence of two signals in the ferromagnetic phase, Paramagnetic signals fit in to a single Lorentzian. The evolution signals with the temperature and their fittings are shown in the figure 9. The continuous black line shows the experimental data and the red line shows the fitted data in figure 9. EMR spectral parameters like central field (Ho), full width at half maximum (FWHM) and integrated intensity are extracted by fitting them to the suitable equation at all the temperatures except at the phase transition region. The variation of these parameters with the temperature for NSMO-8 and for NSMO-11 are shown in the figures 10 and 11.

These parameters show the changes in the trend at the respective Curie temperatures. On the basis of the relative magnitudes of parameters and their temperature dependence, we assign the high field signal (HFS) to the spins in the shell region and the low field signal (LFS) to the spins in the core region. HFS and LFS are shown in red and black symbols as shown in fig 10,11.It can be seen that LFS is considerably more intense than the HFS corresponding to a much larger number of spins in the core region than in the shell region. Moreover, as reported in the literature [8], the core region is ferromagnetically ordered where as the shell spins are either weakly magnetic or magnetically disordered. Therfore, the core spins are subject to the Weiss field , which gets added to the applied exteranal field making their resonance appear at lower field. This interpretation is also consistent with the widths of the two signals: The core spins , being ferromagnetically ordered are subject to shape dependent demagnetization fields. From this one expects them to have larger linewidths as indeed observed. Very interestingly, this behaviour of linewidth of the signals from the two regions is exactly opposite to that in the bulk. In bulk manganite samples, one observes two FMR signals ,though for an entirely different reason viz, phase segregation [26,27,30]. In these samples, the high field signals have larger linewidths than the low field signals. In contrast, we observe in nanomanganites that the low field signals have larger linewidths for the reasons mentioned above. This provides further support to the assignment of the signals. NMR studies have shown that a nano grain consists of core and shell regions, which are different from one another in magnetic nature. The NMR signal which is coming from the core is attributed to $Mn^{3+/4+}$ ions and the core signal is due to $Mn^{+4}$ only[12,25]. By comparing the intensities of two signals at T < 180K, where the intensities are nearly temperature independent, one can estimate the relative number of spins in the core and shell regions. We find that for the nano particles of size ~20 nm (radius ~ 10 nm), the shell radius is of the order of 0.85 nm to 1.2 nm. This is consistent with the report of Bibes et.al [11] who estimated a shell thickness of ~1nm in their LCMO nanoparticles. Interestingly, it is seen that the g-value in the paramagnetic phase increases (from 1.9806-1.9852) with the decrease of particle size i.e) spin-orbit coupling and crystal field anisotropy are affected by the particle size which are known from the literature. Linewidth magnitude (gives information about the spin dynamics) also changes with the particle size in the paramagnetic phase. The observed variation of EMR spectral parameters of 15nm and 30nm particles may be due to the variation in the domain nature i.e) 15nm particles are of single domain and 30nm particles are having both single and multidomains[7]. Experimental and theretical investigations are in progress to know the further insights.

**Conclusions:** Fine particles of perovskite manganite NSMO have been prepared sucsessfully by sol-gel method of size 30-50nm. These are characterized by XRD, TEM and EDAX. The sol-gel prepared samples show very different magnetic and transport behaviour compared to that of their bulk counterpart of this composition. We also able to report the variation in EMR spectral properties with temperature of different particle sizes. The core-shell model is experimentally probed using the EMR technique. The shell thickness is estimated from the EMR spectral parameters is 0.7-1nm and is in agreement with the existing theoretical and experimental estimates.

**Figure Captions:**

1. XRD pattern of NSMO-8, NSMO-11.
2. a) TEM micrographs of NSMO-8

   b) TEM micrographs of NSMO-11. Scale bar is in micrometer.
3. A typical plot of resistivity versus absolute temperature of NSMO-8 at different magnetic fields.
4. A typical plot of real part of the AC magnetic suceptibility ($\chi$) versus absolute temperature (T) for NSMO-9 sample. Inset plot is its first derivative ($d\chi/dT$).
5. Variation of crystallite size, $T_p$ and $T_c$ with varying temperature

6. Resistivity versus absolute temperature curve for NSMO samples below $T_p$ both in presence and in the absence of magnetic field.
7. Variation of $Ln(\rho/T)$ as a function of inverse temperature (1/T), for (a) NSMO-8, (b) NSMO-9, (c) NSMO-10 and (d) NSMO-11 above $T_p$ both in presence and in absence of magnetic field.
8. Plot of $\ln\sigma$ versus $T^{-1/4}$ for (a) NSMO-8 in the absence of magnetic field.
9. Typical EMR signals at different temperatures for NSMO-8.
10. Variation of EMR spectral parameters central field (Ho), full width at half maxima (FWHM) and normalized intensity (NormI) with the temperature for NSMO-8.
11. Variation of EMR spectral parameters Ho, FWHM and NormI with temperature for NSMO-11.

**Figures:**

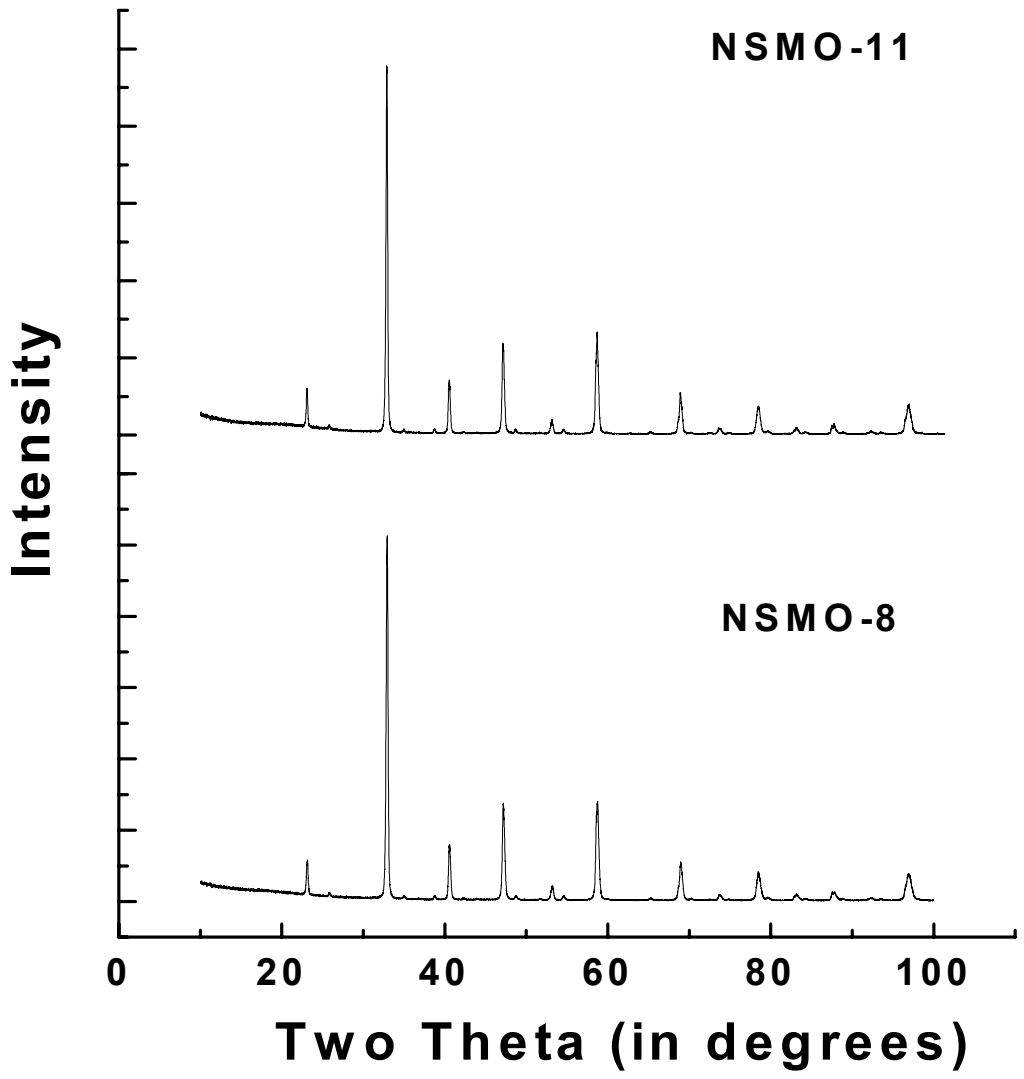

**Fig1**

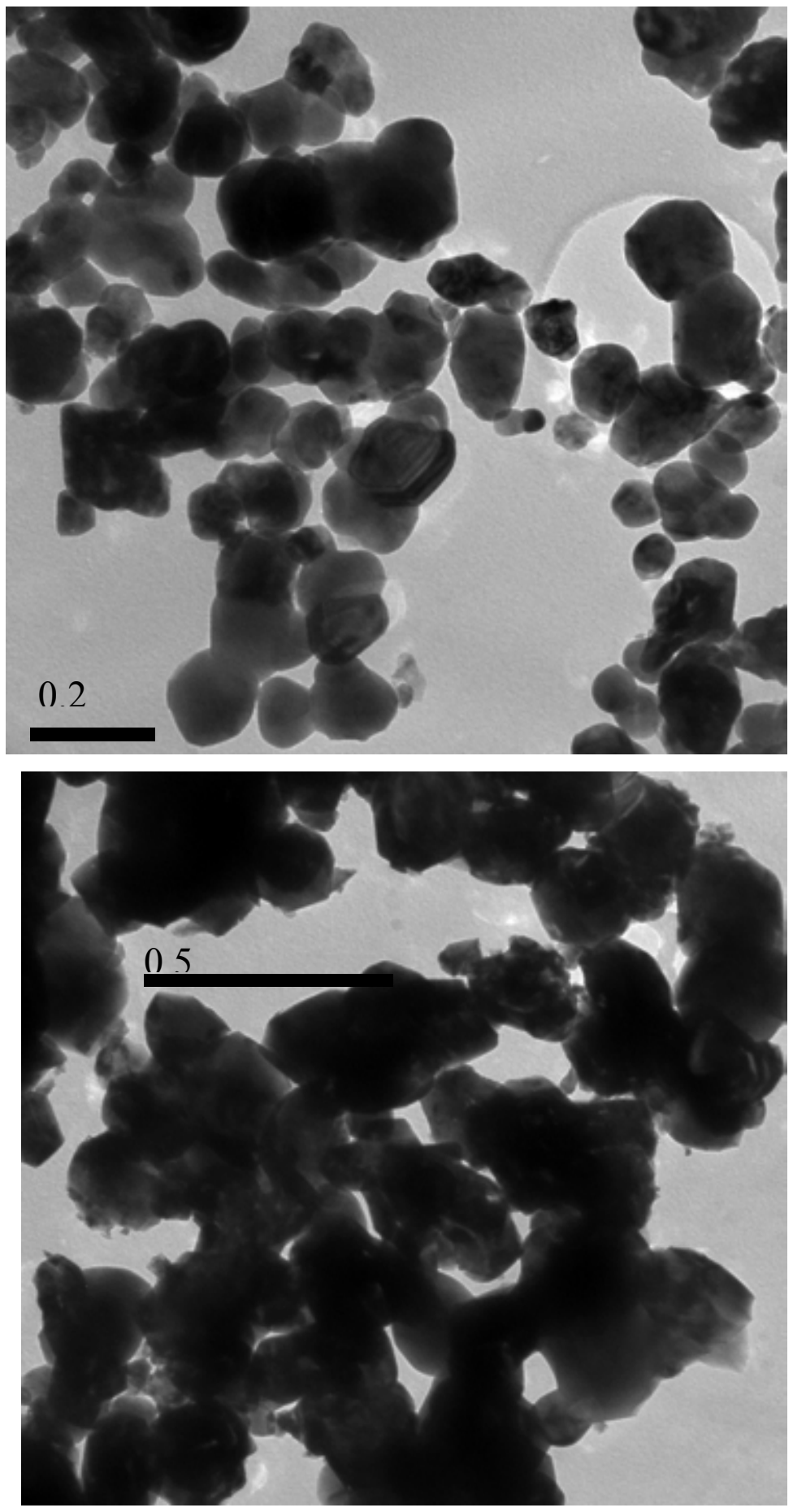

**Fig 2(a), 2(b)**

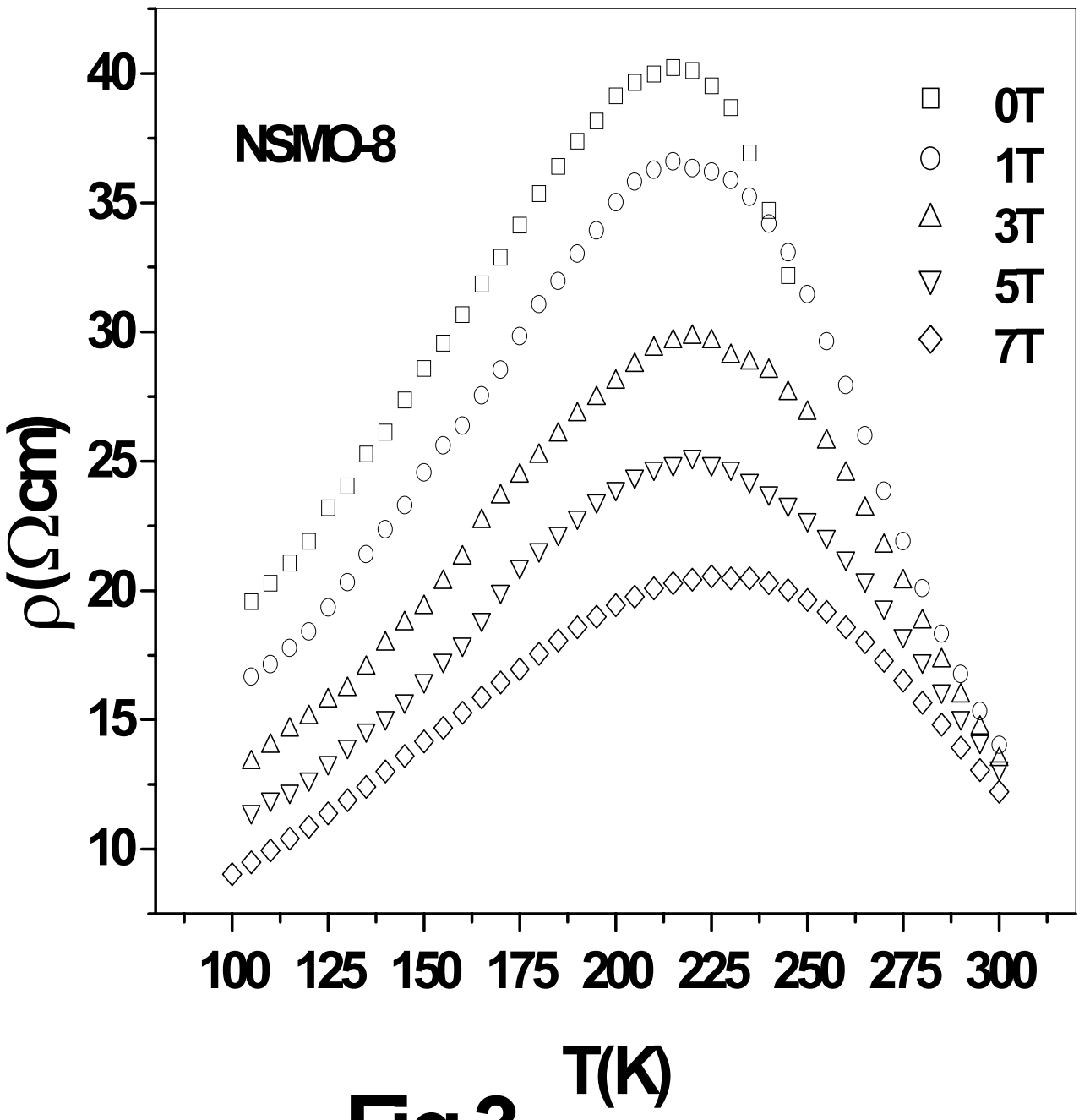

Fig 3

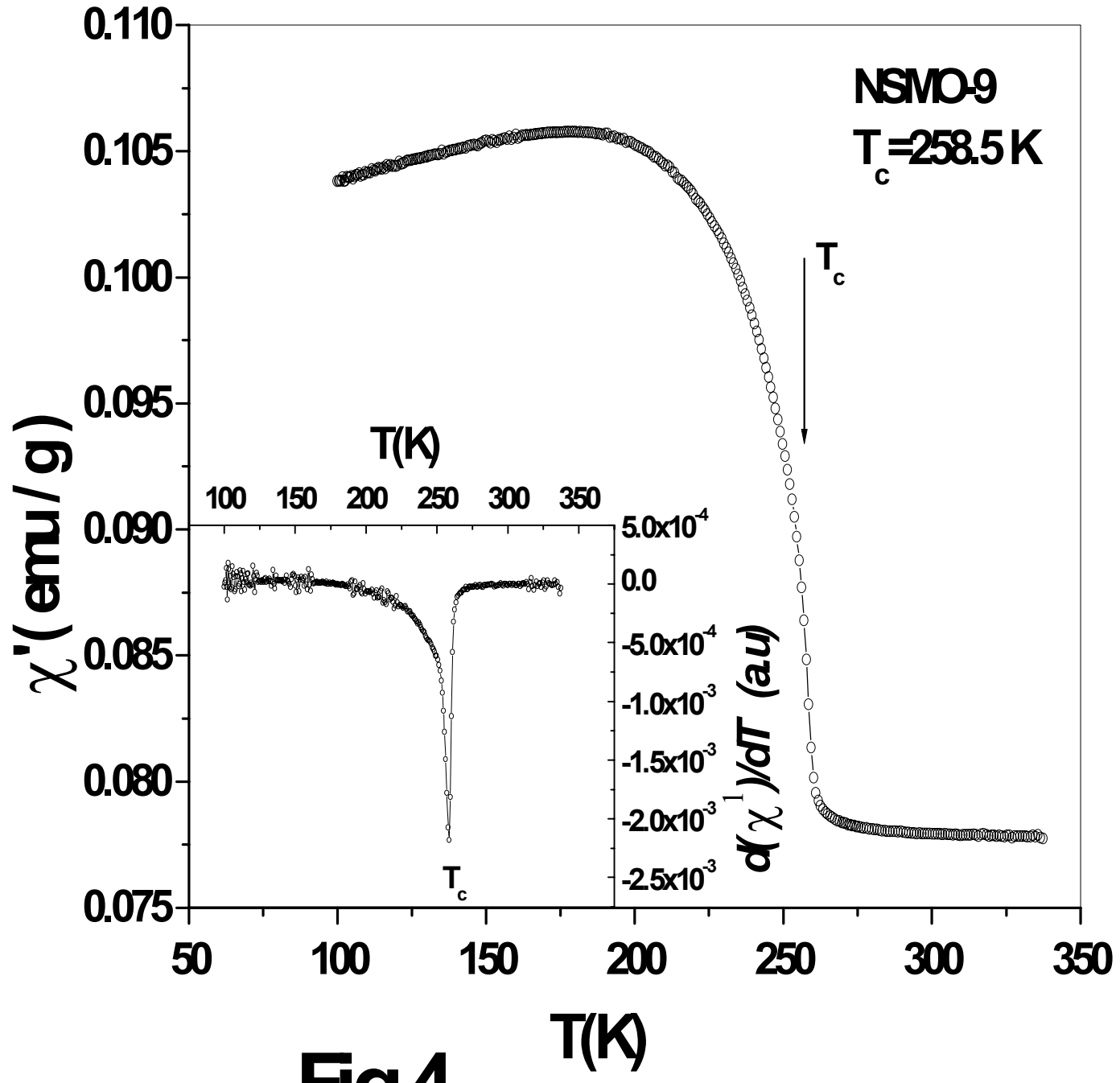

Fig 4

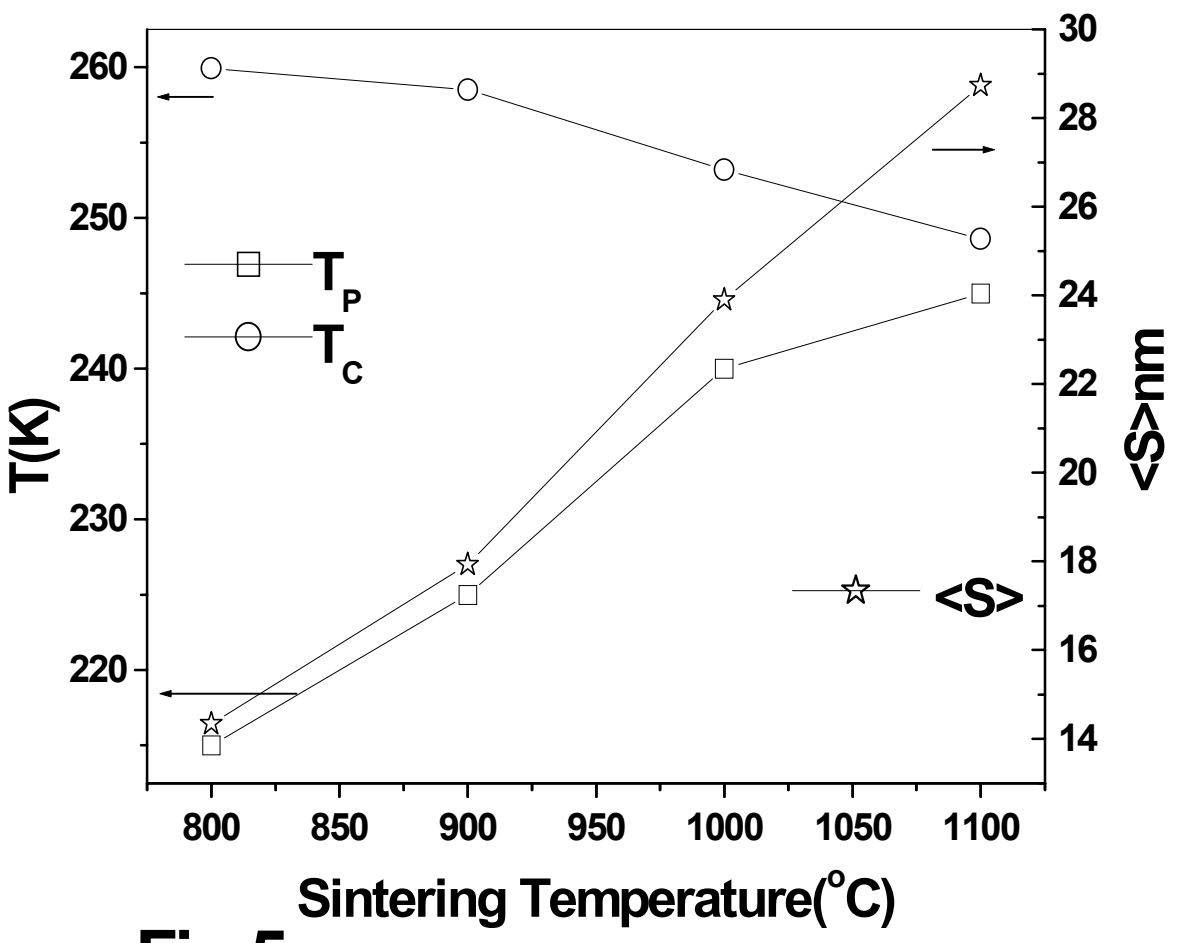

Fig 5

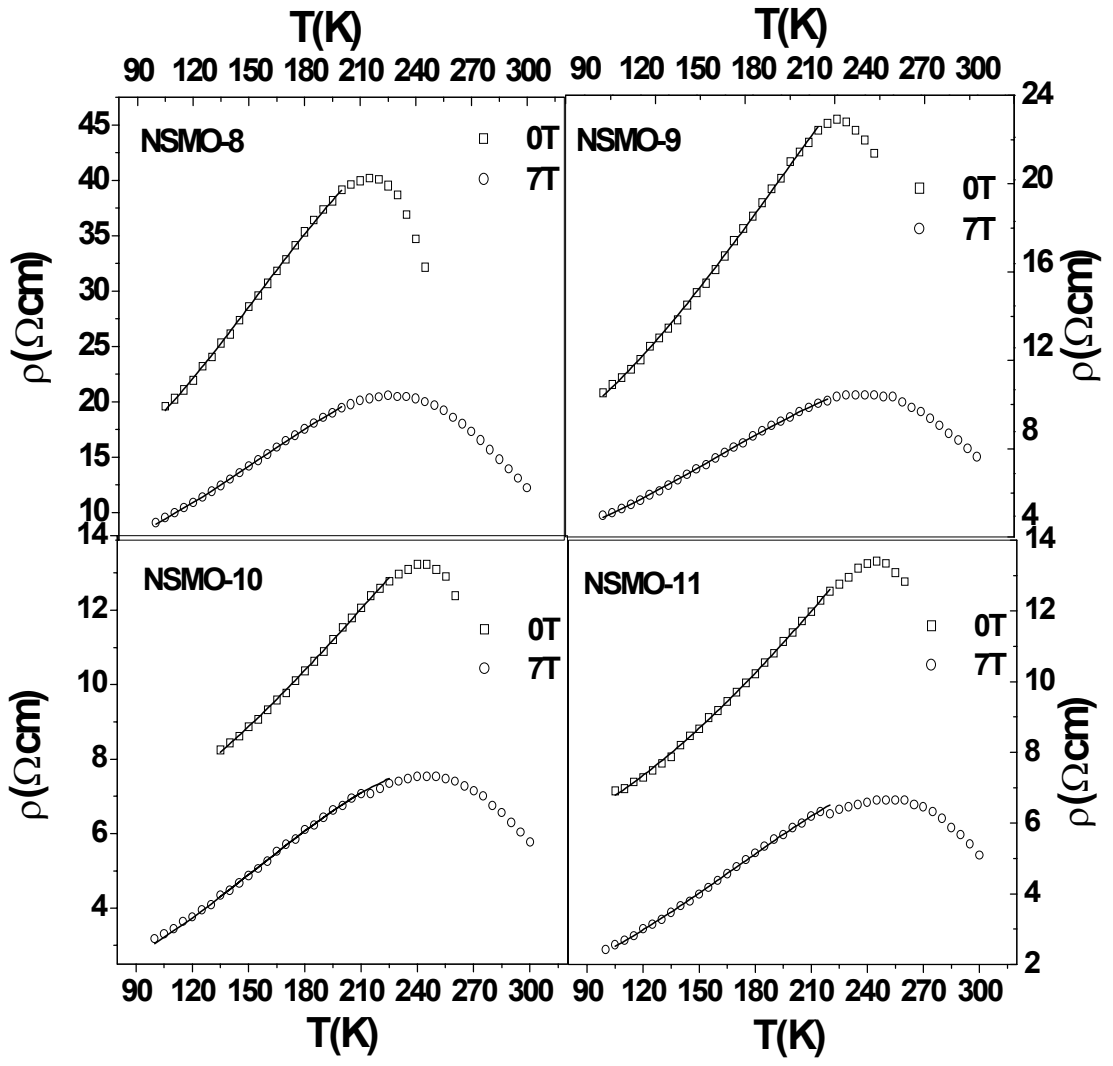

Fig 6

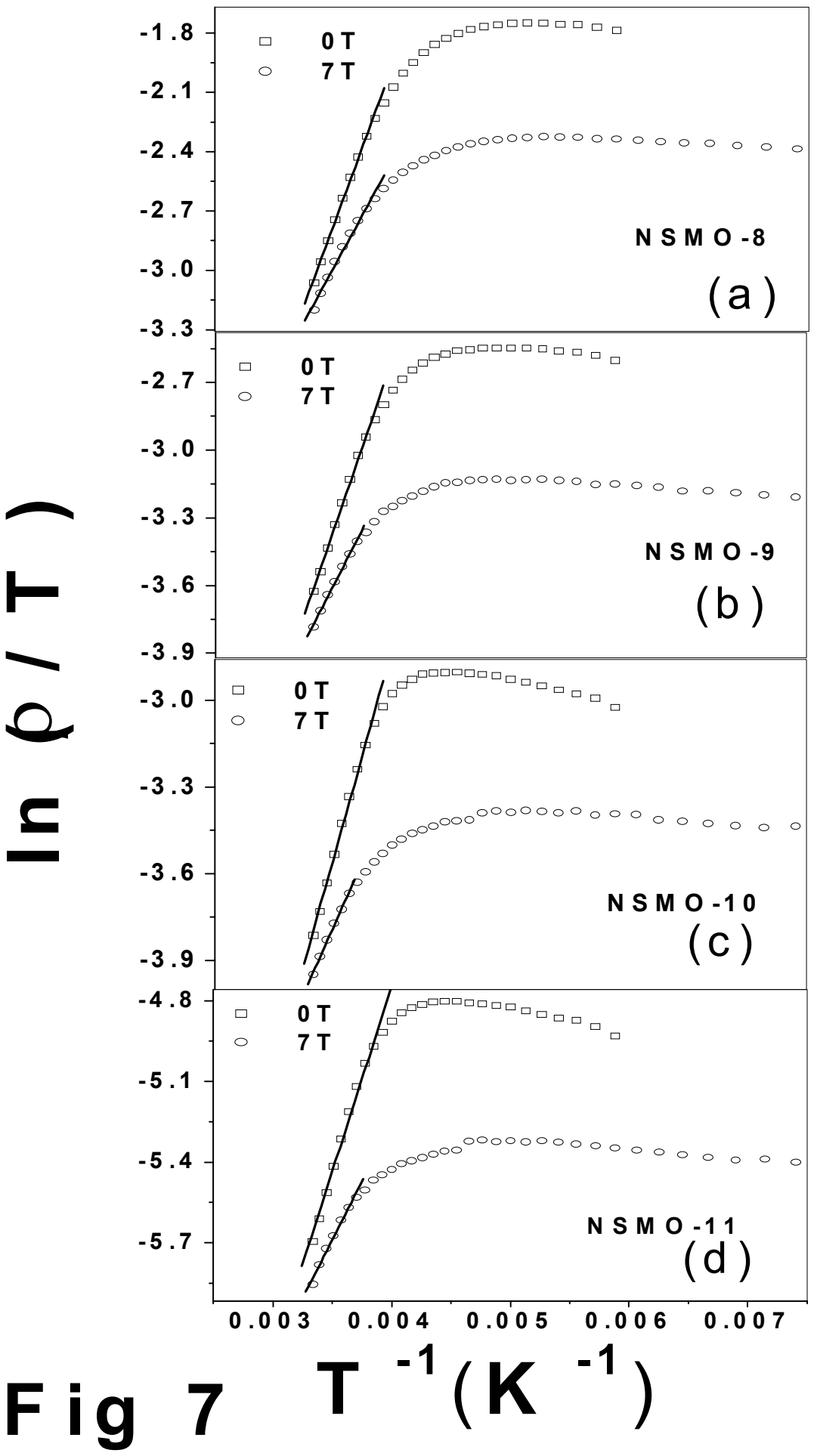

Fig 7

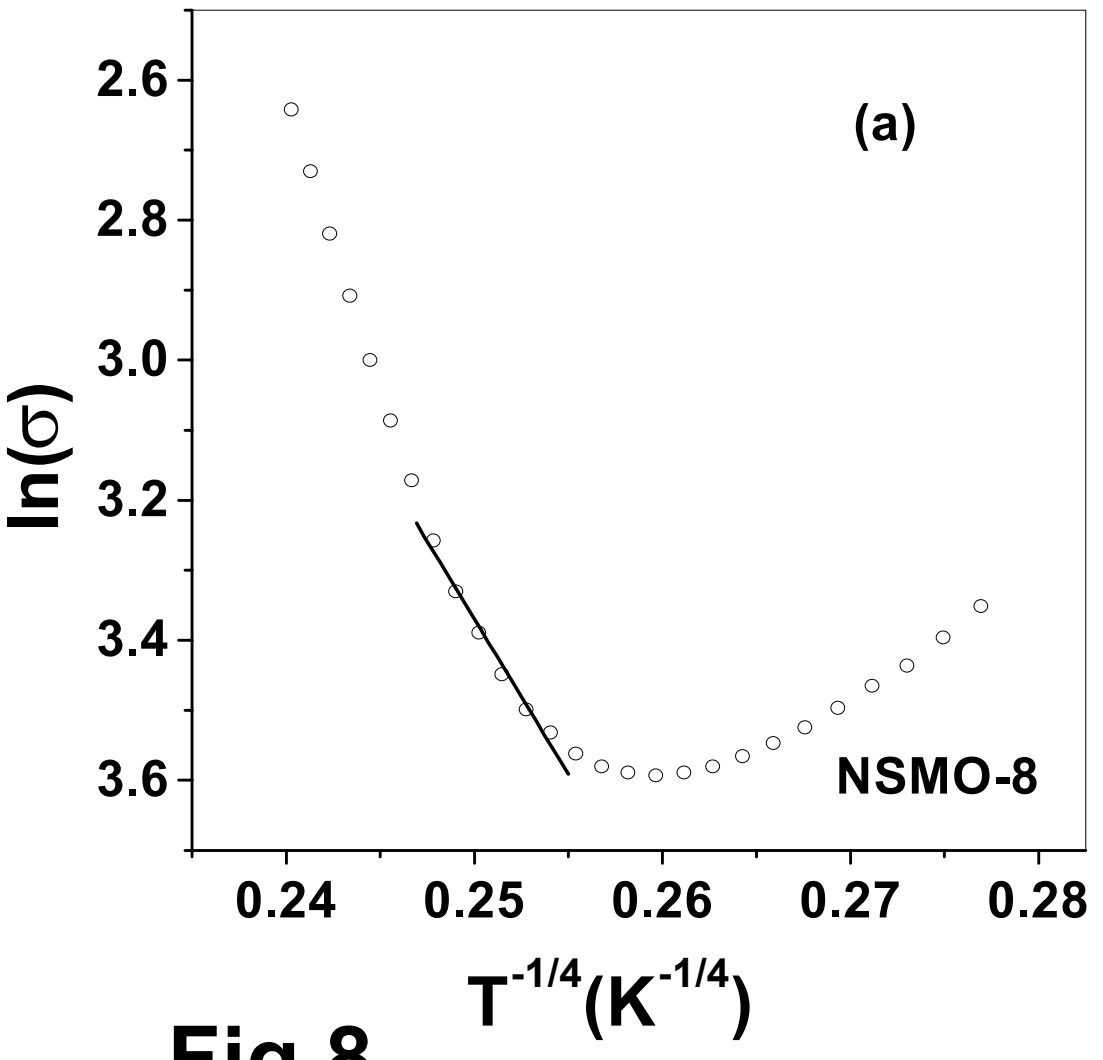

Fig 8

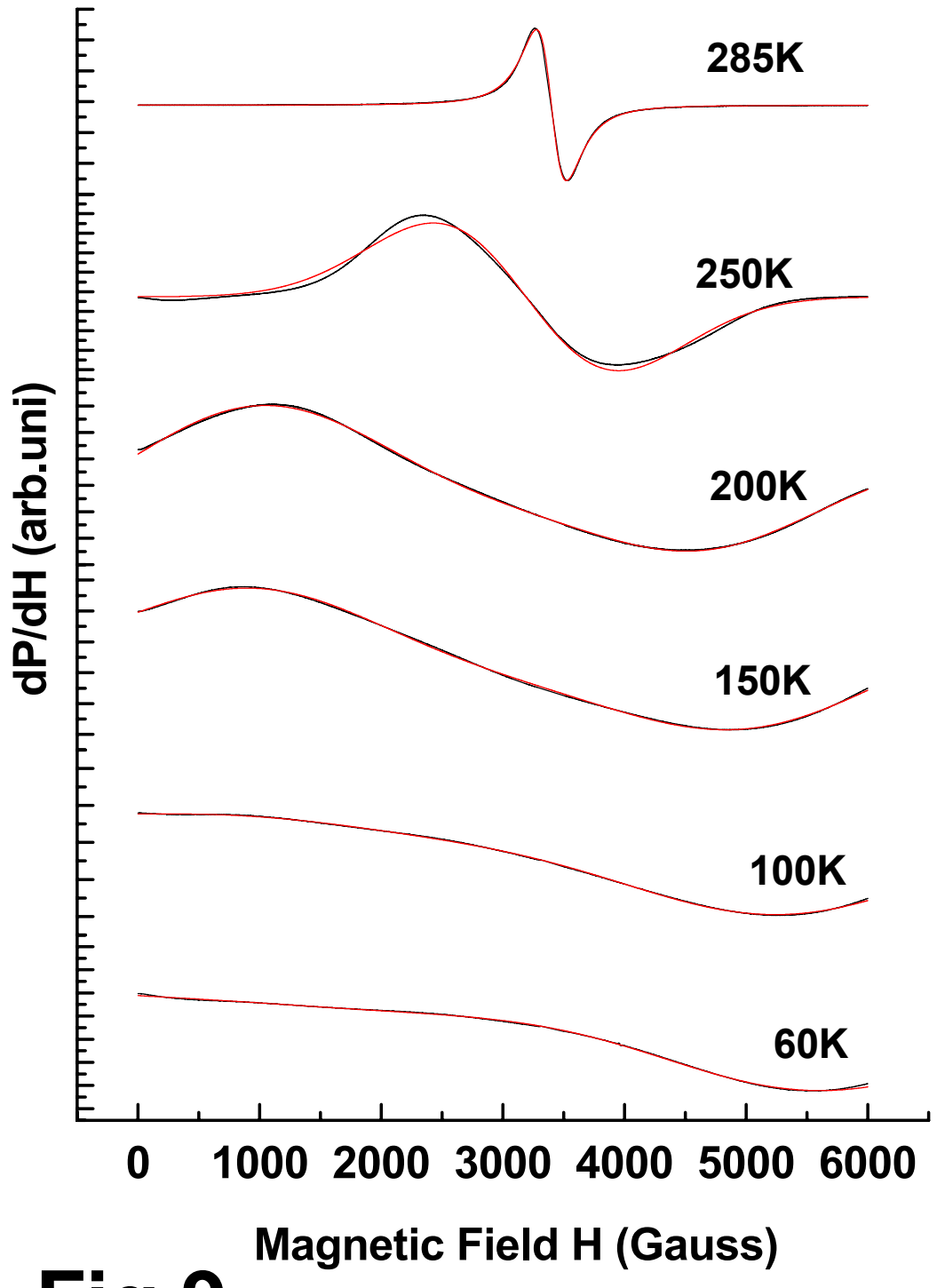

Fig 9

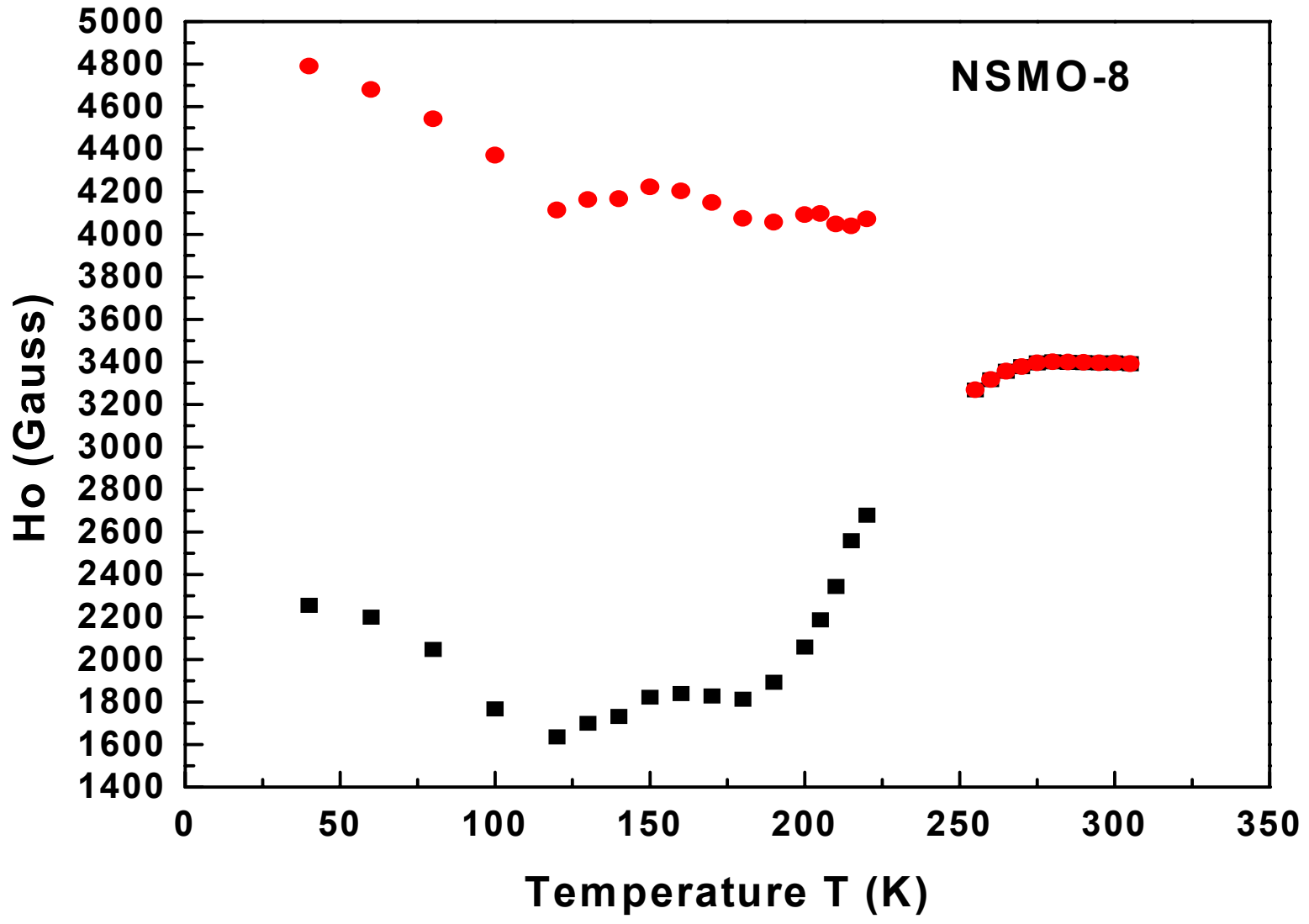

Fig 10 (a)

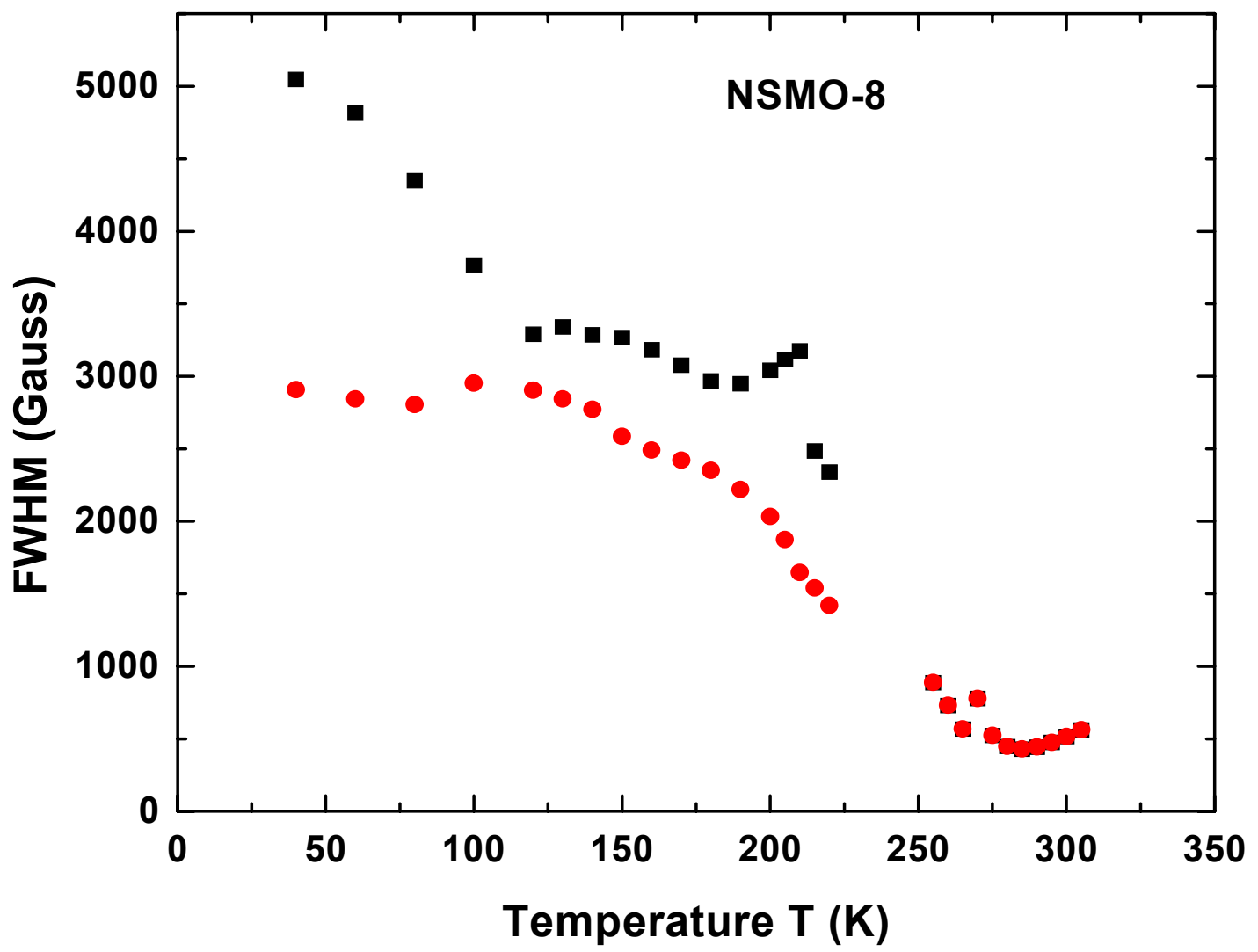

Fig 10 (b)

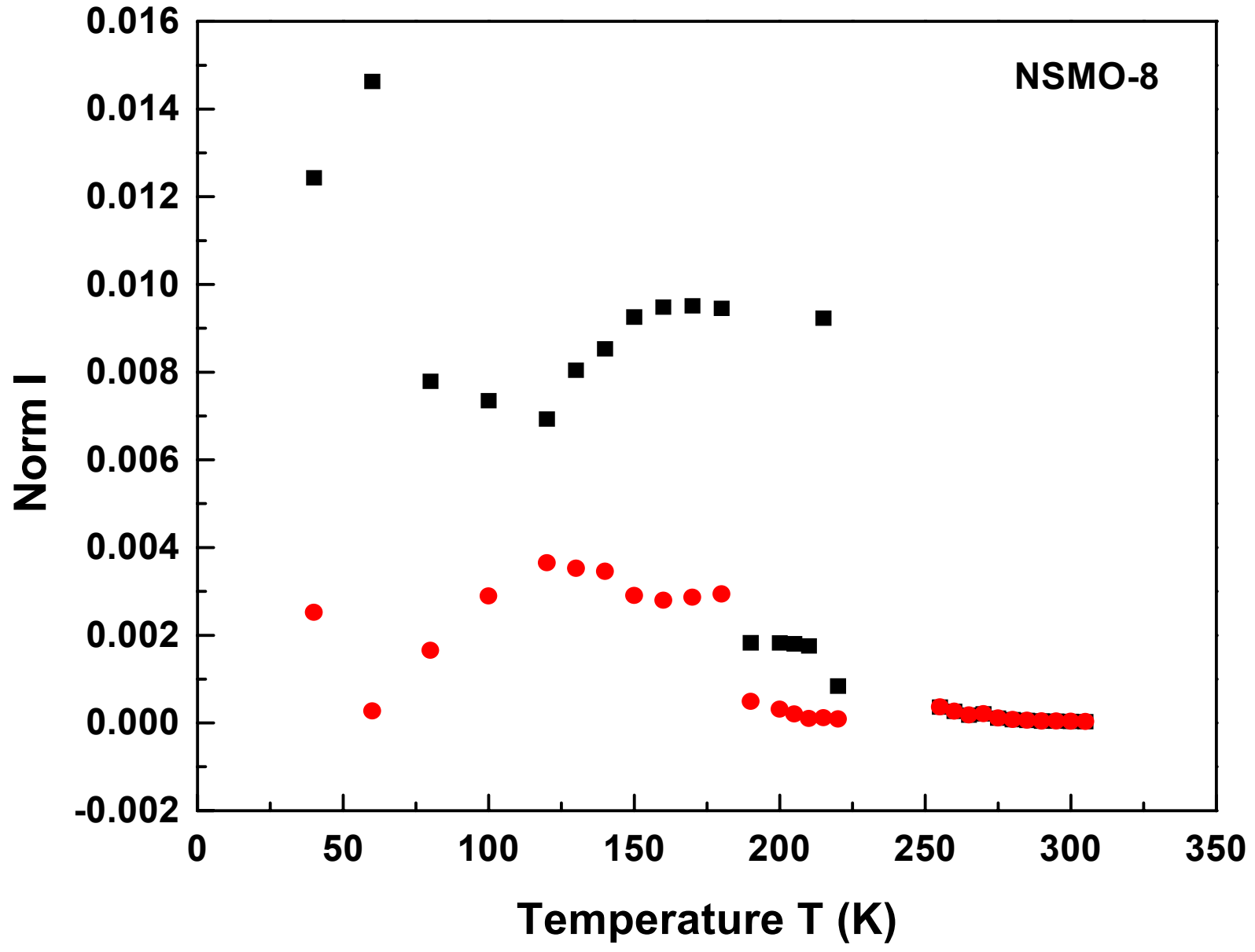

Fig 10 (c)

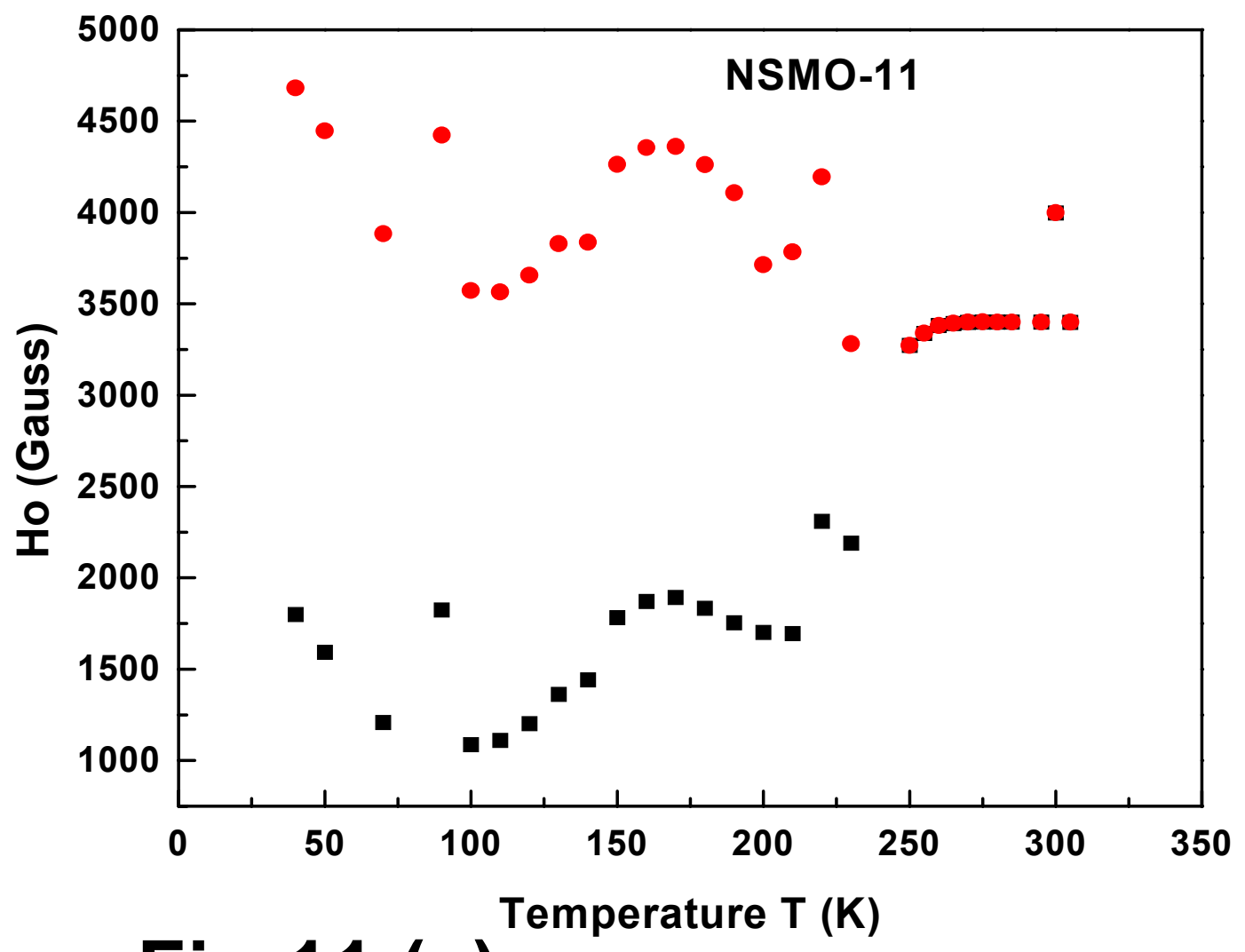

Fig 11 (a)

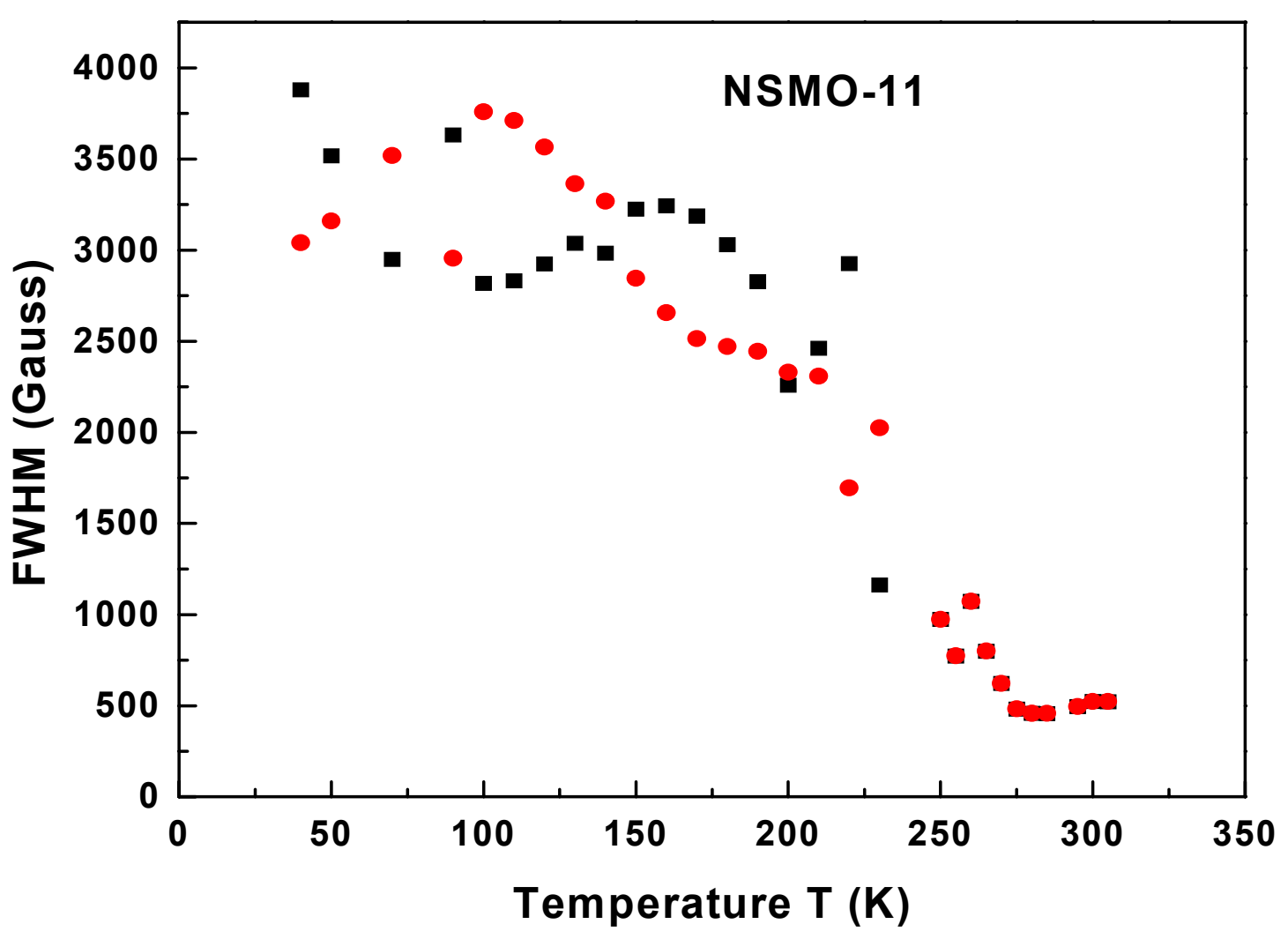

Fig 11 (b)

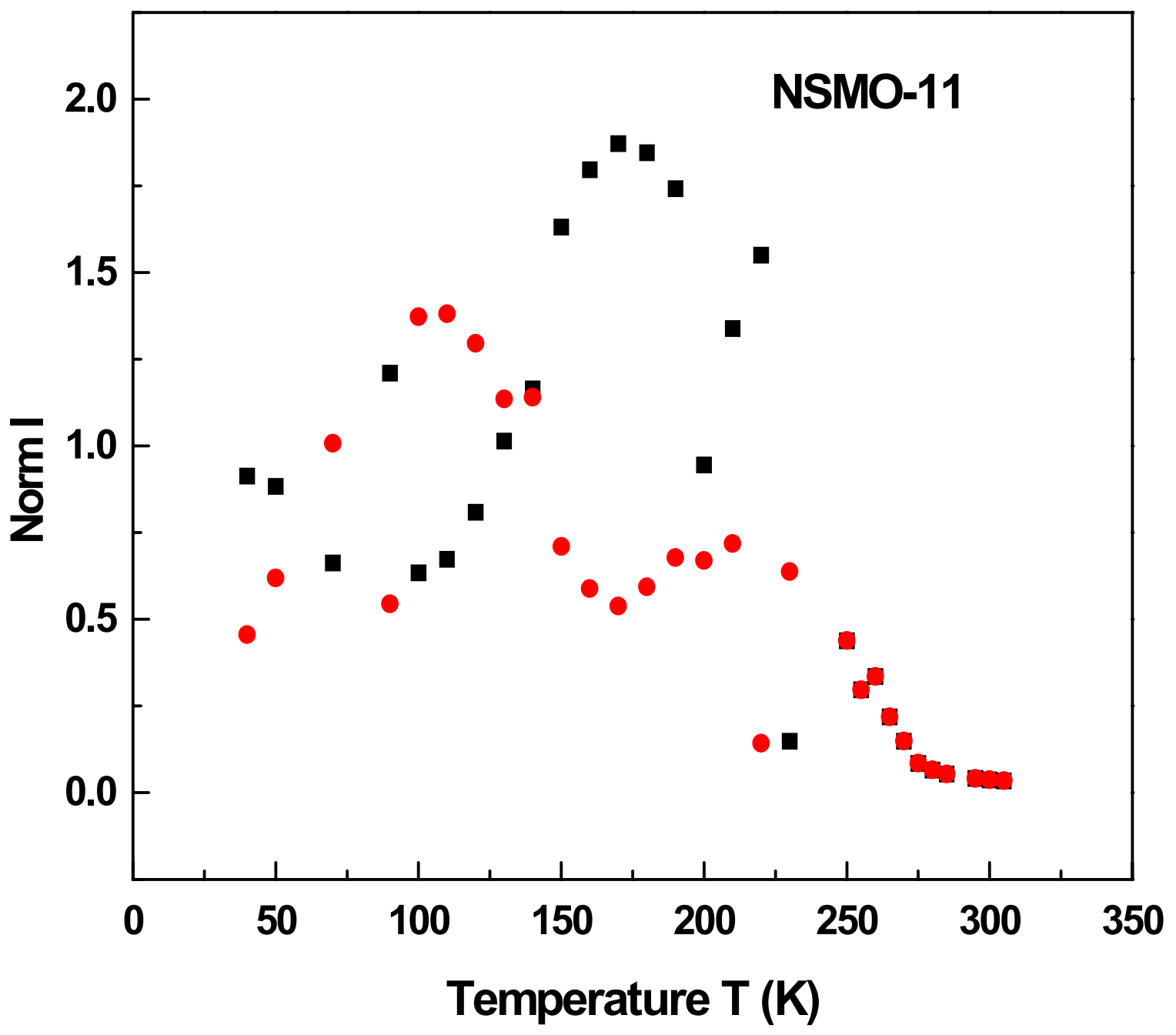

Fig 11 (c)